\documentclass[linenumbers,preprint,superscriptaddress]{revtex4}
\usepackage{graphicx}

\begin{document}

\title{Similarity analysis of macroecology}

\author{S. C. Chapman}
\affiliation{Physics Department, University of Warwick, CV4 7AL, UK}
\author{N. W. Watkins}
\affiliation{Physics Department, University of Warwick, CV4 7AL, UK}
\affiliation{London School of Economics and Political Science, London, UK}
\affiliation{British Antarctic Survey, NERC, High Cross, Madingley Road, Cambridge CB3 0ET, UK}
\author{G. Rowlands}
\affiliation{Physics Department, University of Warwick, CV4 7AL, UK}
\author{A. Clarke}
\affiliation{British Antarctic Survey, NERC, High Cross, Madingley Road, Cambridge CB3 0ET, UK}
\author{E. J. Murphy}
\affiliation{British Antarctic Survey, NERC, High Cross, Madingley Road, Cambridge CB3 0ET, UK}




\begin{abstract}
We perform a full similarity analysis of an idealized ecosystem using Buckingham's $\Pi$ theorem to obtain dimensionless similarity parameters given that some (non- unique) method exists that can differentiate different functional groups of individuals within an ecosystem. We then obtain the relationship between the similarity parameters under the
assumptions of (i) that the ecosystem is in a dynamically balanced steady state and (ii) that these functional groups are connected to each other by the flow of resource. The expression that we obtain relates
the level of complexity that the ecosystem can support to intrinsic macroscopic variables such as density, diversity and characteristic length scales for foraging or dispersal, and extrinsic macroscopic variables such as habitat size and the rate of supply of resource. This expression relates these macroscopic variables to each other,   generating commonly observed   macroecological patterns; these broad trends simply reflect the similarity property of ecosystems. We thus find that  details of the ecosystem function are not required to obtain these broad  macroecological patterns this may explain why they are ubiquitous.  Departures from our relationship may indicate that the ecosystem is in a state of rapid change, i.e., abundance or diversity explosion or collapse.
Our result provides normalised variables that can be used to isolate the trend in one ecosystem variable from another,    providing a new method for isolating macroecological patterns in data.
A dimensionless control parameter for ecosystem complexity  emerges from our analysis and this
    will be a control parameter in dynamical models for ecosystems based on energy flow and conservation and will order the emergent behaviour of these models.
\end{abstract}


\maketitle

\section{Introduction}
The rapid increase in the availability of large-scale ecological data \cite{Brown1995,Maurer1999} has increased  knowledge of global patterns and stimulated the search for the underlying processes that determine them (see \cite{Marquet2009}).  Examples of the large-scale, macroecological, patterns \cite{Rosenzweig1995} to have emerged from empirical analyses include the species-area and the species-latitude relationships, and trends in density and diversity with body size.  These are broad scale patterns and generalised rules rather than mechanistic processes, and a range of theories have been proposed to explain why these patterns emerge.  This has led to the development of a perspective in which the detailed biological characteristics of the species (traits) do not determine their abundance and that the processes affecting community structure can be considered as neutral \cite{Hubbell2001,Harte2008,mcgill2010}, see also \cite{Chave2004,Leibold2004}. The fact that these patterns are both approximate and ubiquitous, suggests that they do not reflect the details of how ecosystems operate, rather that they emerge from underlying general constraints.

The diversity seen throughout ecology has meant that the very possibility of existence of universal laws remains  controversial \citep{Lawton1999,Harte2002,Dodds2009}. In stark contrast,  universal laws are an essential feature of physics. One reason for this dichotomy is the covariance principle, which asserts that the laws of physics can be represented in a form equally valid for all observers \cite{Buckingham1914,Buckingham1915,Barenblatt1996,Bolster2011}. Any physical principles relevant to a classical natural system thus cannot depend on arbitrarily chosen units \cite{Stephens1993} i.e. they must be expressible in terms of   dimensionless parameters which are invariant under a change of fundamental units of measurement. It implies that the functions describing the behaviour of a system   must have arguments which are dimensionless quantities, known as ``similarity parameters" $\Pi_i$ \citep{Barenblatt1996,Barenblatt2002}. Underpinning even this is a deeper classical measurement postulate, that of the possibility of controllable, repeatable, and observer-independent measurement.
Furthermore in physics the existence of symmetries gives rise to conservation principles governing the relationship between physical quantities.  Such conservation properties will   constrain the relationships which can exist between the $\Pi_i$, and will therefore give rise to patterns in these quantities. Ecological patterns  are seen and measured in a range of quantities, some of which have physical units. A natural question \cite{Rosen1989,May2007}, that we address here, then arises: to what extent do physical processes determine the major trends observed in macroecology?

\begin{figure}
\begin{centering}
\includegraphics[scale=0.6]{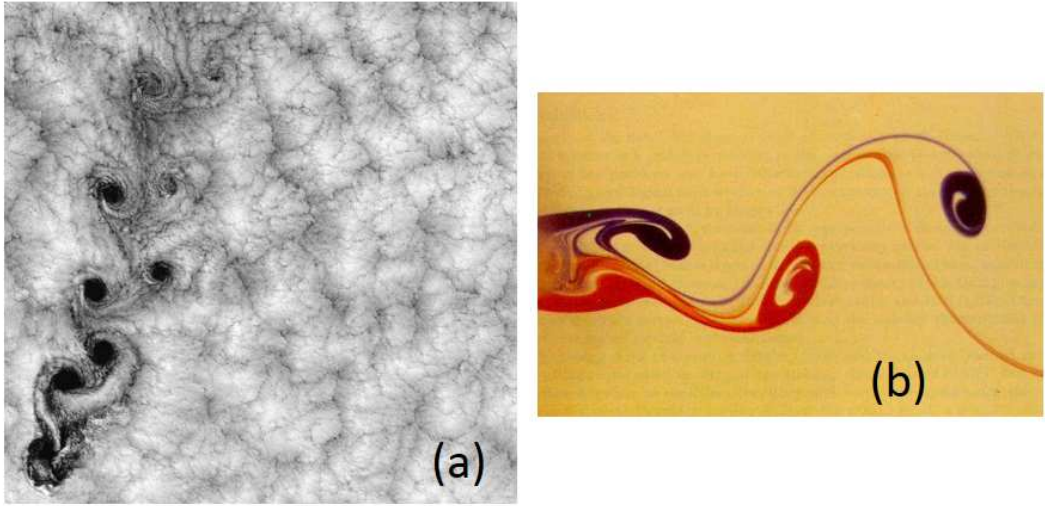}
\end{centering}
\caption{Two systems with the same similarity properties. Von Karman vortex street seen in (a) Landsat 7 satellite image of an island in the clouds and (b) laboratory fluid experiment.}
\end{figure}

We now briefly review the essential idea of similarity analysis with reference to a well known example in fluid dynamics. Figure 1 shows two realizations of a von Karman vortex ``street". These are on quite different scales, they are (a) cloud bearing airflow over an island at sea and (b) a laboratory based experiment in water (for experiments of this type, see \cite{buff}). The full dynamics is strongly non-linear and the functional form of the observed vortex street pattern is non-trivial to obtain. Nevertheless, we can see that these patterns are quite similar and we can characterize the vortex street by two lengthscales, the width of the pattern, or the obstacle around which the fluid is flowing ($L_o$), and the distance downstream between successive vortices ($L_d$). Figure 1 (a) and (b) look similar because  the ratio of these lengthscales, $L_o/L_d$ is similar, despite the fact that $L_o$ is tens of kilometers for the island, and is millimeters to centimeters in size in the laboratory. The ratio $L_o/L_d$ is just the dimensionless  Strouhal number, $S=f L_o/U$ in fluid dynamics, since $L_d=U/f$ where $U$ is the background flow speed and $f$ is the frequency at which vortices are shed at the obstacle (in air, the ``aeolian tone"). Clearly, $L_o$ and $L_d$, (or $L_0,U$ and $f$) are governing parameters that characterize the observed vortex street pattern that is formed, with a corresponding dimensionless similarity parameter that is just the Strouhal number. The question that arises is, what determines this observed pattern, specifically, why are these two situations at very different spatial scales, similar? Now, there are other governing parameters for fluid flow. Let us consider one other, the kinematic viscosity $\nu$ (physical dimension, (length)$^2$/time). A dimensionless parameter involving $\nu$, is $R_e=UL_o/\nu$, the Reynolds number. If there are no other governing parameters, then the observed vortex street pattern, specifically, its Strouhal number, can only depend upon the Reynolds number, that is, $S=F(R_e)$ \cite{Ray}. This result can be tested experimentally by plotting $S$ versus $R_e$. Such a plot reveals that the Strouhal number has only weak dependence on the Reynolds number for $R_e\sim 10^3-10^6$, for which $S\sim 0.2$. This is approximately what we see on Figure 1 where the Reynolds numbers are at either end of this range.

Rayleigh \cite{Ray} used similarity analysis to obtain the similarity parameters $R_e$ and $S$ for the vortex street without obtaining a model or mathematical form for the detailed nonlinear function that describes them. The origin of the precise functional relationship between $R_e$ and $S$ ($S=F(R_e)$) remains a topic of current research \cite{SR1,SR2}. Notably, Rayleigh's procedure \cite{Rott} was to (i) identify the similarity parameters,  $R_e$ and $S$ then (ii) note that these parameters ordered the experimental data and finally (iii)  propose an approximate form for $F$ that modeled the observed dependence of $S$ upon $R_e$. An important corollary of Rayleigh's result is that the observations, when plotted in \emph{dimensional} units, will show patterns or trends, for example, the frequency $f$ will increase linearly with flow speed  $U$, for a given size obstacle. Furthermore, if data from a number of experiments with different obstacle size are aggregated, such a plot will still show this trend, but with considerable scatter.  Any organizing principle based on physical properties will thus only clearly emerge if the dataset is plotted in terms of dimensionless similarity parameters.
In this paper we will perform this important first step, we will obtain similarity parameters for ecosystems. This will suggest a new method to test this data for such organizing principles, and we suggest may explain observed macroecological patterns or trends.

 Importantly, similarity does not presuppose  {\it self}-similarity but does encompass it.   Self-similarity is the property of generalized power law dependence where there is no characteristic scale,
  (e.g  \cite{Barenblatt1996} pp 86ff; \cite{Embrechts}).
 Scaling relationships between subsets of observed macroecological variables have been proposed (see e.g. \cite{Enquist2003,brownrev}) and self-similarity is intrinsic to some models of metabolism, dispersal and foraging \cite{ViswanathanBook}.  Although an important topic, self-similarity is an extra assumption that we do not need to make here, and are not making.

Although most familiar in physics and engineering \cite{Bolster2011}, the application of dimensional techniques to biological problems dates back to Galileo's application of similarity arguments to the load bearing capacity of bones \citep{Thompson1915}. Dimensional arguments find widespread application in the scaling of numerical simulations in biology, as elsewhere.
By contrast, as noted by \cite{Rosen1989} and more recently \cite{May2007}, only very limited explorations have been made so far of the use of  full similarity analysis to codify the set of possible relationships between macroscopic variables in ecology. An interesting exception was the conjecture \cite{lovelock} that ''life-like processes [might] require a flux of energy above some minimal value in order to get going and keep going".  Lovelock made an explicit analogy with the observation by "Reynolds … that turbulent eddies in gases and liquids could only form if the rate of flow was above some critical value in relation to local conditions."

In this paper we will perform such a similarity analysis of macroecological variables.  Rather than exploiting statistical constraints as in  maximum entropy inference \cite{Pueyo2007,Harte2008,Banavar2010,Haeg2010}, we focus on \emph{physical and dimensional} constraints and how these in turn constrain the relationships that can exist between
intrinsic variables such as density, diversity and characteristic lengthscales for foraging or dispersal, and extrinsic variables such as habitat size and the rate of supply of resource.

The starting point in any similarity analysis is to identify the 'governing parameters' \cite{Barenblatt1996}, that is, the variables or quantities that are necessary for any description of the system. Similarity analysis in essence takes these governing parameters as its input, and as its output identifies the dimensionless similarity parameters. Here, we will use the extensive body of observations in macroecology to specify a set of governing parameters. Importantly, these parameters encompass both size based and species based approaches \cite{Keller} to describing an ecosystem.  They are intrinsic variables such as the diversity of species (richness), the density of individuals, characteristic length scales for foraging or dispersal, metabolic rate, and extrinsic variables such as habitat size and the rate of supply of resource. Provided that the governing parameters have physical dimension (dimensional units), similarity analysis constrains the possible relationships that can exist between them. The functional relationships we obtain here are between dimensional quantities, which in practice tend to arise from approaches which focus on flows of materials and energy through the ecosystem.


Similarity analysis does not specify a unique relationship between these parameters.  We can only obtain this if in addition there are known \textit{physical} constraints, such as conservation principles.
 The flow of energy and resource has been one of several key threads of ecosystem analysis \cite{hagen} since for example the work of Lotka, Lindeman and the Odums \cite{L1,L2,L3}.
 The physical constraint that is central to this paper is that we consider ecosystems which adapt dynamically to changes in external parameters to maintain a balance between the rate of uptake and of utilization of resources taken over the ecosystem as a whole. This is our key assumption. This does not mean that the system is in a fixed state, only that over the time scales being considered the inputs/outputs are balanced and this balance constrains the overall structure and functioning of the ecosystem. The concept of ecosystems which compensate dynamically to remain in  homeostasis has been explored previously \cite{ernest2001,white2004} but not developed using dimensional analysis.

In this paper we adopt the formal definition of an ecosystem; that the different functional groups of individuals within the ecosystem are connected to each other by the flow of resource. All individuals within the ecosystem ultimately derive resource from that taken up by that ecosystem's primary producers. We assume that the ecosystem is in a dynamically balanced steady state in that the total rate of uptake of resource is just balanced by the rate summed over the ecosystem at which it is utilized. We will show that this is sufficient to obtain
an expression that constrains the relationships that can exist between these (dimensional) ecosystem variables.  We will see that this constraint is  reflected in overall macroecological trends that are observed. Further, we obtain the relationship between these trends. At minimum this determines for the first time the dimensionless, or normalised variables that isolate trends in one ecosystem variable from another. This new method for isolating macroecological trends provides the basis for understanding dependencies between factors such as size and metabolic rate that quantify the flow of resource, and factors that categorize individuals into distinct species and types. The more resource that is available, the more complex an ecosystem can in principle be, in the sense that more distinct species and types, and relationships between them, can be supported.
At maximum, we obtain a dimensionless control parameter and relate it to this level of complexity that the ecosystem can support.
 This points towards 'thresholds for life', that is, a parametrization of the minimum level of complexity that can potentially be supported by an ecosystem in dynamic balance.

Process based models of ecosystems that aim to predict macroecological patterns tend to fall into two approaches. The first of these relates area, diversity and abundance by means of models for dispersal, occupancy and coexistence/competition  and speciation in physical space (see e.g. \cite{mcgill2010,Marquet2009,Ritchie2010,ODwyer}). The second relates abundance, body mass and metabolic rate by means of models and constraints for the availability and flow of resource (\cite{Enquist2003}, see the review of \cite{brownrev}). The relationship that we obtain here links the key parameters of both these approaches, suggesting a synthesis of them.
It is important to emphasize at the outset  that we do not present a theory of how ecosystems work, nor do we develop an ecosystem model.  Rather we use formal similarity analysis to explore the constraints that must operate on the variables observed in real ecosystems, given the single key assumption of balance between the ecosystem summed rates of resource uptake and utilization.

As noted above many of the specific  relationships between variables that we examine are known and been explored \cite{Rosenzweig1995,Hubbell2001,Harte2008,mcgill2010}.  These include  the species-area, species-latitude and productivity-diversity relationships.  The major insight we provide is to show how these different relationships are related and under what conditions they will emerge.  We also show that they are an expected consequence of the physical constraints on the system.

\section{Similarity analysis and a bottom up-approach to an ecosystem}

The $\Pi$ theorem formalises the principle of similitude as follows. Any physical system that depends upon  $V$  variables, (the governing parameters),  $Q_{1}, Q_2,..Q_V$ will have a function $F$ that relates them: $F(Q_{1}, Q_2,..Q_V)=C$ where $C$ is some dimensionless constant \cite{Barenblatt1996}. The physical system that we discuss here is that which captures general physical aspects of ecosystem function, specifically the uptake and utilization of resource.
The essential idea of similarity analysis is that
this function can only depend upon dimensionless similarity parameters $\Pi_1,\Pi_2,..\Pi_M$ so that $F=F(\Pi_1,\Pi_2,..\Pi_M)$ only. These  dimensionless parameters $\Pi_{1..M}(Q_{1..V})$ are  formed directly from the governing parameters $Q_{1}, Q_2,..Q_V$.
Similarity analysis as in Buckingham's $\Pi$  theorem (\cite{Buckingham1914,Buckingham1915,Barenblatt1996}, see also \cite{Chapman2009}) is simply the process to
obtain the similarity parameters, that is,  these dimensionless groups of variables.
If one then has additional information about the system, such as a conservation property, the  $\Pi_{1..M}(Q_{1..V})$ can be related to each other to make $F$  explicit.
If the $V$ governing parameters $Q_{1..V}$  are expressed in $W$ physical dimensions (i.e. mass, length, time) then from \cite{Buckingham1914} there will be at least $M=V-W$ dimensionless similarity parameters or groups  $\Pi_{1..M}(Q_{1..V})$ which we now progressively identify.

We will build our understanding by first considering the simplest possible idealized ecosystem and then successively increasing the level of complexity; at each stage, additional governing parameters (ecosystem variables) are introduced. Our approach is to use similarity analysis at each stage to find the constraints that act on this general description of an ecosystem. Importantly, we seek to describe an 'observed macroscopic ecosystem', that is, the variables that we will ultimately identify include observed intrinsic properties such as density, diversity, body size, and metabolic rate.
 Introducing progressively more specific detail inevitably introduces more governing parameters- this procedure could be taken further to explore specific detailed ecosystem models by the input of more detailed ecosystem functional properties into the dimensional analysis. Our aim here is rather to explore the ecosystem constraints that emerge for the minimum set of assumptions and model inputs.

\subsection{One kind of uniformly distributed single cell organism.}

\begin{figure}
\begin{centering}
\includegraphics[scale=0.6]{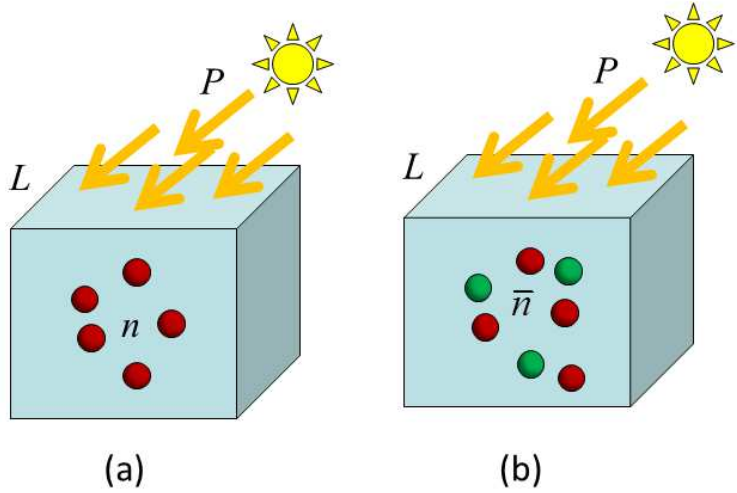}
\end{centering}
\caption{Bottom up ecosystem (a) only one type of single cellular organism can be distinguished (b) 2 types of single cellular
 organism can be distinguished.}
\end{figure}

The ecosystem is composed entirely of single cell organisms that are of uniform type: they have the same function and structure and same typical metabolic rate $R$, dimensions of power $[M][L]^2 [T]^{-3}$. They are uniformly distributed in a habitat of size $L$ in $D$ dimensions with density $n$, dimensions $[L]^{-D}$. The available resource (sunlight) is delivered at rate $P$ per unit area, dimensions power per unit area $[M][T]^{-3}$ and the
(dimensionless) fraction $\alpha$  taken up averaged over the ecosystem is a constant. A schematic of such a system is shown in Figure 2 (a). We recognize that such a simple system does not exist in reality.  However, this simple theoretical construct is an important first step in the analysis, from where we can move to consider more realistic ecological scenarios.

There are  5 governing parameters ($R,L,P,\alpha,n$) and 3 physical dimensions (mass $[M]$, length $[L]$, time $[T]$) so we have 2 similarity parameters (dimensionless groups) which are
\begin{equation}\label{PIset1}
\Pi_1=\frac{\alpha P L^2}{R},\Pi_2=nL^D
\end{equation}
These are related by the physical constraint that the system is in dynamical balance so that the rate at which energy is taken up over the ecosystem is the rate at which it is utilized, so that:
\begin{equation}\label{simple}
\frac{\alpha PL^2}{R}=nL^D
\end{equation}
which is  $\Pi_1=\Pi_2$. This expression simply tracks the flow of energy into and through the ecosystem- it assumes that all other processes necessary for the ecosystem to function, such as the recycling of resources such as Nitrogen,  occur. Introducing a typical metabolic rate for the single cell organisms has fixed
 an energetic minimum threshold for life which is $\Pi_1=1$, that is,  one cell in the habitat (one cell ecosystem).

\subsection{ More than one kind of single cell organism.}

We next consider single cell organisms that can be differentiated from each other, either by their function or their structure, or both. These would represent distinct types or species and different methods for categorizing and distinguishing individuals will yield different sets of species. What will follow will be independent of the precise details of this differentiation method, we only need assume that such a differentiation is now possible. A schematic of such a system is shown in Figure 2 (b). In the ecosystem there are $S$ types (species) and there is a species label $k=1..S$, the density of the $k^{th}$ species is $n(k)$.
We can always define an average density of the single cell organisms:
\begin{equation}
\bar{n}=<n(k)>_k=\frac{1}{S}\sum^{S}_{k=1}n(k)
\end{equation}
so that the variable $n$ in (\ref{PIset1}) is now replaced by $\bar{n}$ and $\Pi_2=\bar{n}L^D$.
The additional variable $S$ is dimensionless; so that we now have 6 governing parameters and 3 dimensions, and
so 3 dimensionless similarity parameters with $\Pi_3=S$. These are again related by the physical constraint that the system is in dynamical balance:
\begin{equation}\label{constraint1}
\frac{\alpha PL^2}{R}=\sum_{k=1}^S n(k)L^D =\bar{n} S L^D
\end{equation}

Expression (\ref{constraint1}) now encapsulates the idea of  primary producers- one or more of the species is responsible for the uptake of resource with efficiency $\alpha$. The other species 'feed off' this primary producer either by grazing, predation or uptake of waste.

\subsection{Multicellular organisms}

We now consider more complex organisms that are multicellular. All the organisms live in an ecosystem and are connected to the primary producers by the flow of resource, either directly or indirectly by predation, or both. It is now possible to distinguish types of organism  and we will label the different types or categories distinguished in this way with index $p$.
Again, the results to follow will not depend upon the precise details of how individuals are assigned to any of the $p$ categories, simply that such an assignment can be made.
 Different methods for categorizing the individuals in an ecosystem \cite{Gotelli2001} will organize individuals into groups or categories of different $p$, this categorization may focus on the functional role of individuals in the ecosystem such as niche or trophic level, or may focus on stage of development, or other factors. Organisms falling into a given $p^{th}$ category or group will be clustered around an  average body size, on average they will be composed of  $B(p)$ cells (this is typically observed \cite{Rosenzweig1995}), and will have average metabolic rate $RB(p)$, the per-cell metabolic rate $R$ now corresponds to an  ecosystem average over these multicellular organisms.
There is a non- trivial correspondence between average size $B(p)$ and how complex an organism can be. Within each $p$ there will be a number $k=1..S(p)$ of differentiable species each with density $n(k,p)$ all with average body size $B(p)$ and with average density, for that $p$ of
\begin{equation}
\bar{n}(p)=<n(k,p)>_k=\frac{1}{S(p)}\sum^{S(p)}_{k=1}n(k,p)
\end{equation}

Our governing parameters, density and diversity,  are now taken to relate to the observation of a given category $p_*$ that is embedded in the ecosystem; we observe $n_*=\bar{n}(p_*)$, $S_*=S(p_*)$.  Individuals in the observed category also have a characteristic average size $B_*=B(p_*)$ (number of cells so dimensionless) which introduces an additional governing parameter giving a total of 7
governing parameters and hence 4 dimensionless similarity parameters:
\begin{equation}\label{PIset2}
\Pi_1=\frac{\alpha P L^2}{R},\Pi_2=n_*L^D,\Pi_3=S_*, \Pi_4=B_*
\end{equation}

The physical constraint of a dynamically balanced ecosystem is now:
\begin{equation}\label{bal1}
\frac{\alpha P L^2}{R}=\sum_{p} \sum^{S(p)}_{k=1}n(k,p) B(p)L^D=\sum_{p}\bar{n}(p)S(p)B(p)L^D
\end{equation}
We can write (\ref{bal1}) in terms of our observed $p_*^{th}$ category:
\begin{equation}\label{bal2}
\frac{\alpha P L^2}{R}=  n_*S_*B_* L^D\sum_{p}\frac{\bar{n}(p)S(p)B(p)}{\bar{n}(p_*)S(p_*)B(p_*)}
\end{equation}
We then have an expression of the form:
\begin{equation}\label{bal3}
\frac{\alpha P L^2}{R}=  n_*S_*B_* L^D \Psi(p_*)
\end{equation}
where $\Psi(p_*)$ is a dimensionless function; $1/(\Psi(p_*))$ is the fraction of the total rate of resource supplied to the ecosystem that is utilized by the observed ($p_*^{th}$)  category. Importantly, all of the species and categories are connected into the same resource flow, so that fundamentally, (\ref{bal3}) will constrain how observed density, diversity and body size are related to each other across the ecosystem, with consequences for macroecological patterns as we will discuss.

These dimensionless similarity parameters (\ref{PIset2}), and their relationship (\ref{bal3}) express the following fundamental properties of the idealized ecosystem. There is a building block on which life is organized- the single cell which has a definable typical (ecosystem average) metabolic rate, $R$.  There is then the physical property of resource flow: that the single cell metabolic rate, along with the rate of uptake of resource to the ecosystem $\alpha P L^2$ constrains the number of cells the ecosystem can support, which is $\Pi_1$. The detailed biological and ecological properties of the ecosystem then organise these $\Pi_1$ cells into a complex network of species and groups of species, observationally these are characterized into functional units which have an average body size, density and diversity.
Hence, the observed ecological variables are in a macroscopic sense related to each other by the physical property of resource flow.

\subsection{Non uniform distribution of individuals in space.}

Generally, organisms will not be uniformly distributed in space so that the observed density depends on the length scale $r$ over which it is observed, so that $n=n(r,k,p)$ and similarly, the efficiency of the primary producers, which depends on their density, is  $\alpha=\alpha(r)$.
The lengthscale over which the density varies can either arise from how individuals subdivide and grow, forage, or other forms of influence they have on each other and on the environment. This is important since in (\ref{bal1}-\ref{bal3}) the density refers to that measured over the habitat of the entire ecosystem on scale $L$ and any observation will be on a more local scale $r<<L$, which in turn relates to the lengthscale of over which the density varies. This will introduce a variable for the scale of observation of the  $p_*^{th}$ category; $r_*$ (dimension $[L]$) finally giving 8
governing parameters so 5 dimensionless similarity parameters:

\begin{equation}\label{PIset3}
\Pi_1=\frac{\alpha_* P L^2}{R},\Pi_2=n_*L^D,\Pi_3=S_*, \Pi_4=B_*, \Pi_5=\frac{r_*}{L}
\end{equation}
where $\alpha_*=\alpha(r_*)$ so that all variables refer to a consistent set of observations on lengthscale $r_*$.
The density can be generally expressed as $n(r,k,p)=n(L,k,p)/g(k,p,r/L)$ where $g$ is dimensionless and expresses the spatial variation of the $p^{th}$ category; similarly $\alpha(r)=\alpha(L)/g_\alpha(r/L)$.
If the categorization $p$ is based on function and structure then one can anticipate that an average of $g$ over the $S(p)$ species in the category is meaningful so that:
\begin{eqnarray}\nonumber
\bar{n}(r,p)=<n(r,k,p)>_k=\frac{1}{S(p)}\sum^{S(p)}_{k=1}n(r,k,p)\\
=\frac{1}{S(p)}\sum^{S(p)}_{k=1}\frac{n(L,k,p)}{g(k,p,r/L)}
=\bar{n}(L,p)/\bar{g}(p,r/L)
\end{eqnarray}
If we explicitly reference lengthscale $L$,  expression (\ref{bal2}) is:
\begin{equation}\nonumber
\frac{\alpha(L) P L^2}{R}=  n_*(L)S_*(L)B_*(L) L^D\sum_{p}\frac{\bar{n}(p,L)S(p,L)B(p,L)}{\bar{n}(p_*,L)S(p_*,L)B(p_*,L)}
\\\label{bal5}
\end{equation}
for a consistent set of observations for all $p$ categories on the same lengthscale $r_*$ this is:
\begin{equation}
\frac{\alpha(r_*)g_\alpha(r_*/L) P L^2}{R}
=n_*(r_*)g_*(r_*/L)S_*(r_*)B_*(r_*)L^D\Psi(p_*,r_*)
\end{equation}
or writing the spatial variation in a single function $G(r/L)=\bar{g}(r/L)/g_\alpha(r/L)$
\begin{equation}\label{balfin}
\frac{\alpha_* P L^2}{R}= n_*S_*B_*G(\frac{r_*}{L})L^D \Psi
\end{equation}
Spatial variation in density thus leads to spatial trends in diversity, or species- area rules; this has arisen  quite generally as a consequence of the physical constraint (\ref{balfin}) and we will discuss this next.

 Essentially, (\ref{balfin}) is:
\begin{equation}\label{piresult}
\Pi_1=\Pi_2\Pi_3\Pi_4G(\Pi_5)\Psi
\end{equation}
  The dimensionless functions $\Psi$ and $G$ contain all the details of the ecosystem function; $\Psi$ encapsulates the details of how individuals in the ecosystem are categorized and $G$ the details of how these categories of individuals are dispersed in space. Our expressions (\ref{balfin}) and (\ref{piresult}) do not specify a particular model for an ecosystem. Instead, they specify the relationship between available resource, and the level of complexity that the ecosystem can attain.
  The diversity and complexity possible in an ecosystem is constrained physically by the total living biomass within it, and this in turn is constrained by the rate of uptake, and utilization, of resource. We thus identify the ecosystem control parameter $\Pi_1=\alpha P L^2/R$ namely (productivity) $\times$ (habitat size)/(typical metabolic rate) which is just the number of 'typical' cells the ecosystem can support (we can always define an ecosystem averaged metabolic rate per cell $R$). This control parameter constrains the level of complexity that an ecosystem can support in the sense that it constrains the number of different possible configurations, or ways that this total living biomass can be arranged into distinct forms of life.
  For example, if there is only one distinguishable kind of organism in the ecosystem, there is only one  $p$ and $S$ value and $\Psi=1$, we essentially recover (\ref{simple}) where the average metabolic rate of the organisms is $RB_*$. The threshold for one such organism of size $B_*$ to be supported by the ecosystem is $\Pi_1/\Pi_4=1$ or $\alpha_* P L^2/(RB_*)=1$.
 As the ecosystem becomes more complex, $\Psi>1$ and each $p$ category utilizes a smaller share of the total resource supplied. The dimensionless function $\Psi$  thus operates as an order parameter of the ecosystem which reflects the level of complexity.

  Importantly, the function $\Psi$ also incorporates the method of categorization. The complexity of an observed ecosystem inevitably depends in part on how the observed data are categorized. However, the observed values, i.e., the observed density, diversity and so forth of a given category, will depend on how the observer defines that category, i.e. what organisms are included in it. These ideas can be used to re-order the observations to understand the role played by how the data is categorized as we discuss in the appendix.

  A physical  analogy to this is the relationship between the Reynolds number in turbulence and the number of excited modes or degrees of freedom. The Reynolds number is the ratio of a rate of energy input on the largest, driving scale to a rate of energy dissipation on the smallest scale, as is $\Pi_1$ here, and similarity analysis, along with the assumption of steady state (no energy pile up) is sufficient (see eg \cite{Chapman2009}) to constrain  the number of degrees of freedom to grow with increasing Reynolds number.

\section{Constraints on macroecological patterns within and across ecosystems}

Observations both within and across ecosystems consist of specifying a method for classifying  individuals into particular groups or categories and then for each of the $p^{th}$ categories, observing the average density, diversity,  bodysize, and metabolic rate.  From  the constraint (\ref{balfin}):
 we see that these variables are not independent, and (\ref{balfin}) suggests  relationships between them which we will now discuss.

Let us consider that a scheme for classifying individuals is consistently adopted, and observations of
average density, diversity,  bodysize, and metabolic rate of these categories are made. These observations simultaneously collect a range of values of $n_*,S_*,B_*$ for a given $r_*$, $L$ and $P$.
We will first consider the case where 'similar' ecosystems are compared, or where a comparison is made within a single ecosystem, that is, the order parameter $\Psi$ is not varying. Subsets of the variables in equation (\ref{balfin}) will then show functional relationships, this has been found for example by \cite{StorchGaston2004,Storch2005} who demonstrate a relationship between species richness, area and a measure of productivity.

 For 'similar' ecosystems then, expression (\ref{balfin}) is
   \begin{equation}\label{balfin2}
\frac{\alpha_* P L^2}{R}= n_*S_*B_*G(\frac{r_*}{L})L^D
\end{equation}
  This constrains overall patterns or trends, it defines a single multivariate surface in the variable space of density, diversity and so forth. The observed macroecological patterns are paths on this surface. The surface then relates these macroecological patterns to each other through the single expression, equation (\ref{balfin2}). As a first example, let us consider trends that can occur as the available resource over the ecosystem $ \alpha PL^2$ is varied.
 From expression (\ref{balfin2}), one cannot have arbitrary increase in density and diversity with resource, it is constrained. Thus (holding all other variables constant) if resource rate of uptake doubles, and density doubles, diversity cannot increase. If resource is increased by a factor $A$, and the number of species doubles, then the density can only increase by factor $A/2$. This will be the case for any model which has our assumption of a dynamically balanced steady state. This constraint on how ecosystem properties such as density and diversity can vary is our main result. This points to a need to isolate changes in one variable from another and we will provide a method for this.

We can formalize these constraints as follows. Expression (\ref{balfin2}) is
 \begin{equation}\label{balwright}
\alpha_* P L^2= S_*\left[R n_*B_*G(\frac{r_*}{L})L^D\right]
\end{equation}
 An increase in diversity with total net productivity integrated over the habitat is Wright's Rule (the species-energy relationship) \cite{Wright1983}. However, to only see an increase in diversity, one would also need the contents of $[...]$ to be constant, that is, $K$ constant in:
 \begin{equation}\label{balwright2}
 R n_*B_*G(\frac{r_*}{L})L^D=K
 \end{equation}
Thus if a set of observations are indeed across an ecosystem, in the sense that all the observed categories are linked by the flow of resource, then when Wright's rule is seen, the 'resource flow constraint' (\ref{balwright2}) should also be seen.

The general relationship that we have derived (\ref{balfin2}) presents, for the first time as far as we are aware, a view of how the major physical and biological variables that determine key aspects of the structure and functioning of ecosystems are related.   It relates the scale and energy turnover to the variables that describe the internal structure of the ecosystem (the complexity).  The relationship emerges as a result of the dimensions of the underlying variables, but is also constrains and gives the wider context for the relationships between specific variables.  It shows that although specific variables are related, the relationship is dependent on the modifying effects of other interacting variables.  The variables are not independent but instead can co-vary and equation (\ref{balfin2}) provides the basis for understanding how they are expected to interactively affect the relationships between specific variables.     This is a crucial point as it highlights why particular relationships can emerge only under particular conditions.  In the following we consider a set of well known macro-ecological relationships to show how they emerge from our analysis and also what equation (\ref{balfin2}) tells us about how the relationship is affected by the other key variables.
 For each of these macro-ecological relationships there will be a resource flow constraint in the sense of (\ref{balwright2}) which we will now identify:

\begin{itemize}

 \item  \emph{Diversity and Wright's Rule:} $S_*\propto \alpha PL^2$ as in (\ref{balwright}) so that the number of species (diversity) increases with total net productivity integrated over the habitat rather than productivity alone;
 this is Wright's rule \cite{Wright1983}. Whilst Wright's rule is to some extent ecologically trivial (a greater net energy input allows more individuals, see \cite{ClarkeGaston2006}) the interesting aspect of this result is that it predicts an increase in diversity (richness) and not just individuals. As we would anticipate from the resource flow constraint(\ref{balwright2}) the relationship between productivity and species diversity also varies with spatial scale as is found \cite{Gillman2006}. Equation (\ref{balwright}) also suggests that the internal configuration of the ecosystem in terms of density, biomass and metabolic rate of the species present will affect the relationship.

 \item \emph{Diversity and metabolic rate:} $S_*\propto 1/R$ since:
 \begin{equation}
S_*=\frac{1}{R}\left[\frac{\alpha_* P L^{2-D}}{n_*B_*G(\frac{r_*}{L})}\right]
\end{equation}
so that diversity decreases with increasing  metabolic rate: we expect ecosystems dominated by endothermic  organisms with high metabolic rate to have lower diversity than those dominated by ectothermic, low-metabolism, organisms (e.g. \cite{Rosenzweig1995}). The resource flow constraint is now $n_*B_*G(\frac{r_*}{L})L^{D-2}/(\alpha_* P)=K$ constant, which specifies how variation in the scale and productivity of the ecosystems considered could mask the effects of changing metabolic rate.

\item \emph{Latitudinal gradient rule:}   Diversity will also increase with resource, since:
\begin{equation}\label{lat}
S_*=\frac{\alpha_* P}{\left[R n_* B_*G(\frac{r_*}{L})L^{D-2}\right]}
\end{equation}
The resource flow constraint is now
$R n_*B_*G(\frac{r_*}{L})L^{D-2}=K$ constant.
Provided that other factors, i.e. $\alpha_*$, that link resource uptake to available sunlight do not vary \citep{ClarkeGaston2006} our general macroecological relationship encapsulates the latitudinal gradient rule. Again, this trend is present alongside patterns in the other variables when $K$ is not constant, as discussed by \cite{Hillebrand2004}. For example, (\ref{lat}) predicts that low metabolic rate ecosystems where the rate of resource supply is high will be more diverse that high metabolic rate ecosystems where the rate of resource supply is low; this may suggest a refined version of the latitudinal gradient rule and allow comparison of diverse ecosystems.

\item \emph{Species Area Relationships:} A corollary of length-scale dependence of the density is  that diversity will vary  with habitat size (which is a function of $L$) since:
   \begin{equation}\label{sara}
S_*= \frac{L^{2-D}}{G(\frac{r_*}{L})}\left[\frac{\alpha_* P}{R n_*B_*}\right]
\end{equation}
     More explicitly, diversity will vary both with habitat size (which is a function of $L$) and the lengthscale of the observation or characteristic lengthscale of some process (which is a function of $r_*$); these are known as Species Area Relationships (SAR) (see eg \cite{Dengler2009}).
    Thus if the individuals grow in clumps, say by division, or live on a fractal structure (tree, coral, mountain, river) or forage in a random walk pattern  (\ref{sara}) will constrain the resulting ecosystem SAR.
For example power law SAR  arise if available productive surface area or volume orders the availability and uptake of resource\cite{Haskell} and that this is in turn ordered by the roughness of the terrain which can be modeled simply as a fractal  \cite{Ritchie1999,Palmer2007}, see also \cite{Milne1992}. The resource flow constraint is $\alpha_* P/(R n_*B_*)=K$ constant.
 Thus these SAR and the underlying constraint on dispersal and clumping from which they originate are also  found to interact with  other variables such as productivity \cite{Chia2006} and bodysize \cite{Etienne2004}.
Observations over the largest regional or continental scales  tend to integrate or aggregate over detailed spatial dependence  and over other variable factors such as metabolic rate. These scales exceed that over which the terrain varies, and over which processes occur that yield spatial clumping, over such large scales the effective $G \rightarrow 1$ and the landscape is essentially 'flat' so that $D=2$. Hence on these largest scales the spatial dependence  (\ref{sara}) vanishes, we have that $P \sim S_*$, and
 positive relationships between productivity and diversity emerge as has been found  \cite{Gillman2006}. Within a given ecosystem, diversity  and abundance will also both vary with the $r_*$ over which they are observed, as well as with each other. If all other variables are not strongly varying, their functional dependence on $r_*$ can be obtained from the data by the method of scaling collapse as has been done by \cite{Zillio}.

\item \emph{Abundance, the 'more individuals' hypothesis:} The abundance (density of individuals in each species) increases with productivity integrated over the habitat and decreases with the typical metabolic rate (e.g. \cite{Rosenzweig1995}), since:
    \begin{equation}\label{morin}
n_*=\frac{\alpha_* P}{R}\frac{1}{\left[S_*B_*G(\frac{r_*}{L})L^{D-2}\right]}
\end{equation}
As we increase the total ecosystem energy uptake rate, from equation (\ref{morin}) both the diversity and abundance can increase (for fixed metabolic rate). We will find the above patterns when one effect does not dominate, for example, Wright's rule will not be seen if the increase is entirely in abundance, and not in diversity. The relationship between the number of individuals and the number of species has long intrigued ecologists, and whilst a positive relationship is sometimes assumed (the 'more individuals' hypothesis for the increase in richness with overall abundance: see \cite{ClarkeGaston2006}) the detailed mechanism(s) involved are far from clear.

\item \emph{Abundance and diversity decrease with increasing with body size:} our expression (\ref{balfin2}) gives an inverse relationship between abundance, diversity and bodysize
\begin{equation}
n_*S_*= \frac{1}{B_*}\left[\frac{\alpha_* P}{R G(\frac{r_*}{L})L^{D-2}}\right]
\end{equation}
  provided the resource flow constraint $\alpha_* P/(R G(\frac{r_*}{L})L^{D-2})$ is constant or weakly varying, indeed
the average abundance \cite{Damuth1981,Cohen2002,Schmid2000,Carbone2002,white2007} and diversity \cite{Rosenzweig1995} are found to have statistical trends that decrease with increasing  average body size.

\item \emph{Trends in trophic level and nett productivity:}
the number of trophic levels that the ecosystem can support is seen to increase with total net productivity summed over the habitat and decreases with the typical metabolic rate \cite{Rosenzweig1995}. This follows since we have from (\ref{balfin}):
\begin{equation}\label{troph}
B_* \Psi=\frac{\alpha_* P L^2}{R}\frac{1}{\left[ n_*S_*G(\frac{r_*}{L})L^D\right]}
\end{equation}
Either bodymass relates to trophic level, or the number of trophic levels may be determined by the level of complexity $\Psi$ which  increases with total net productivity summed over the habitat and decreases with the typical metabolic rate. This pattern will be seen provided $n_*S_*G(\frac{r_*}{L})L^D$ is constant.

\end{itemize}

The above relationships (\ref{balwright})-(\ref{troph}) show that well known macroecological relationships emerge from the single expression (\ref{balfin}) that we obtained from formal dimensional analysis.  This provides a clear physical and ecological basis for understanding why particular relationships exist and have been identified in a wide range of different studies of ecological systems \cite{Brown1995}- \cite{Dodds2009}.  However, our analysis goes further to clarify how these different relationships are related together and are affected by changes in other variables that are crucial for determining the structure and functioning of ecosystems.  It demonstrates how the variables co-vary and why important macroecological relationships may emerge only under particular conditions. We thus expect these patterns to emerge most clearly when the corresponding resource flow constraint is slowly varying and this can be tested for in data.
 It also suggests that taking account of that co-variation will allow a much more rigorous basis for analysing available data to test for the existence of particular macroecological relationships and using those relationships to test our understanding of the factors that determine ecosystem structure and function more generally.  In the following section we illustrate the power of this insight by examining how the co-variability can affect the capacity to detect particular relationships.

\begin{figure}
\includegraphics[scale=0.6]{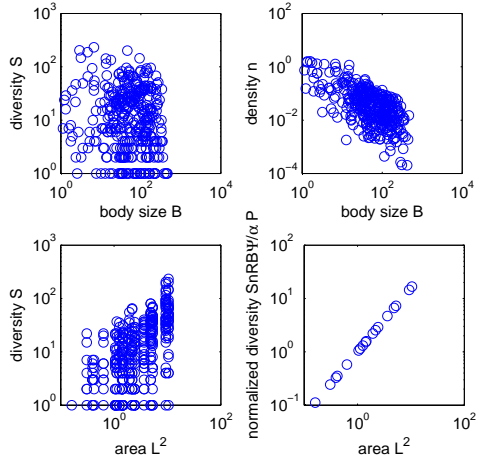}
\caption{Log-log plot of a synthetic dataset generated for  categories of individuals  from habitats of different areas. The dataset is constructed with power law dependence of diversity on lengthscale and trends in abundance, diversity, body size and area with random scatter. This pattern is only revealed in a plot of dimensionless  diversity versus area.
 }
\end{figure}

A corollary of this is that we can use (\ref{balfin}) to identify a method to isolate these patterns. A particular example of this is testing for SAR.
From equation (\ref{sara}), to test for a SAR one should plot the normalized diversity $\bar{S}$ versus $L$:
\begin{equation}\label{bal4}
\bar{S}=\frac{S_*n_*R B_*\Psi(p_*)}{\alpha_* P}=  \frac{L^{2-D}}{G(r_*,L)}
\end{equation}
 Such a comparison can be made if $\Psi(p_*)$ is not strongly varying ('similar' ecosystems), or if
 $\Psi(p_*)$ can be found from the data by the method described in the appendix.

We illustrate this process in Figure 3 where we have modeled synthetic data for a species-area comparison.
We have generated
synthetic data in the same manner as described in the appendix,  such that there are trends in
 abundance, species richness, and body size,
 and also random scatter in all variables, constrained such that all the sampled categories of individuals share the same function $\Psi$. In addition each group of data is from a different habitat size and has a dependence on area predicted by our result (\ref{balfin}) with power law dependence of $G$ on $L$. This simple illustrative exercise demonstrates that an underlying clear pattern emerges in normalized diversity with
 a corresponding
  SAR pattern of diversity versus area which has considerable scatter.

\section{Conclusions}

We have used a 'bottom up' approach to fix the minimum set of governing parameters needed to specify a generic idealized ecosystem and have used these to perform a similarity analysis of the ecosystem. Physical constraints of energy flow and utilization over the ecosystem then relate the similarity parameters, which in turn gives an expression which
 relate the level of complexity that the ecosystem can support to intrinsic variables such as density, diversity and characteristic lengthscales for foraging or dispersal, and extrinsic variables such as habitat size and the rate of supply of resource. These constraints hold regardless of the details of how a given ecosystem functions and require
only  the assumption that the ecosystem is in a dynamically balanced steady state, that is, that the total rate of resource uptake is balanced by the rate of resource utilization summed over the ecosystem.
 We thus find the constraint on the relationships that can exist between these (dimensional) ecosystem variables which    is  reflected in observed macroecological patterns. Our result may explain why these general, approximate  statistical trends appear to be so ubiquitous in nature: we obtain these patterns without recourse to any detailed information about the structure or dynamics of ecosystems or indeed how the data are collected. They simply reflect the underlying similarity properties of the ecosystem and energy conservation in dynamical steady state.

Our result also shows how these different observed macroecological relationships are related to each other and how they are affected by changes in other variables, and hence why particular macroecological relationships may emerge only under particular conditions.
 This leads to the dimensionless, or normalised variables that need to be constructed to isolate the trend in one ecosystem variable from another; we thus provide a new method for isolating macroecological patterns.
  Comparisons could thus be made between datasets by controlling for (normalizing against) characteristic metabolic rate, abundance and diversity in order to isolate the statistical pattern with respect to one of these variables. In particular this method isolates a function that expresses how  complex  the ecosystem is and it would be intriguing to order the data in this way to determine the level of complexity of ecosystems that are found in nature, and to what conditions they correspond.
  An example would be comparisons across extinct ecosystems, or between extinct and contemporary ecosystems, provided a comparable sample group could be identified. The fact that Wright's rule,  species area rules and latitudinal gradient rules emerge often, but not always, from the observational data gathered across ecosystems may reflect varying levels of complexity in these ecosystems, or the effect of different schemes for categorizing individuals within ecosystems
   and  our results provide a method to control for this.

  Departures from these statistical patterns where ecosystems are similar, and consistently sampled,
  then may imply that the system is in a state of rapid change, i.e., abundance or diversity explosion or collapse. Any ecosystem which is dynamically balanced in the sense discussed above will fall within these macroecological patterns, it does not need to be a climax or maximum energy utilization system but simply needs to balance the rate of energy uptake with that of usage integrated over the ecosystem.

    Finally, we have identified a dimensionless control parameter for ecosystem density and diversity, namely (productivity) $\times$ (habitat size)/(typical metabolic rate) which emerges quite generally from our dimensional analysis as a similarity parameter.
   This we suggest will be a control parameter in dynamical models for ecosystems based on energy flow and conservation and will order the emergent behaviour of these models. We relate this control parameter to the level of complexity that a dynamically balanced ecosystem  can support (its order parameter). This control parameter is the ratio of energy input rate to the ecosystem to the metabolic rate of  the smallest possible unit of life, a single cell. If it is reasonable to identify a smallest possible unit of life then this parameterizes the threshold at which life, defined in this manner, can occur (c.f. Lovelock's conjecture about dimensionless control parameters for prebiotic planets \cite{lovelock}).

\section*{Acknowledgments}
The authors acknowledge the UK EPSRC, STFC and NERC for support. This study is part of the British Antarctic Survey Polar Science for Planet Earth Programme. NWW acknowledges valuable interactions with J. E. Lovelock and O. Morton.

\section*{Appendix}

\begin{figure}
\begin{centering}
\includegraphics[scale=0.6]{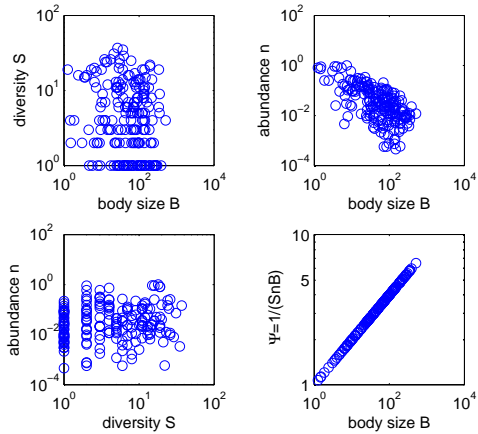}
\end{centering}
\caption{Log-log plots of synthetic data generated for an ecosystem where all the observed individuals are constrained to share the same ecosystem complexity function $\Psi$. The set of data is constructed to show trends in abundance, diversity and body size with random scatter.
The constraint can only be discerned by plotting $1/S_jn_jB_j$ versus body size $B_j$ which to within a constant is the function $\Psi$.}
\end{figure}

Any organizing principle based on physical properties will only clearly emerge if the dataset is plotted in terms of dimensionless similarity parameters. We now demonstrate the procedure to apply this method to macroecological data.
As a starting point say we have a set of observations based on individuals  consistently organized into categories. For each category we observe on lengthscale $r_*$ the density $n_*$, diversity $S_*$ and body size $B_*$, so that for many such categories we have a set of observations of $n_*$, $[n_1,n_2,..n_j...]$, of $S_*$, $[S_1,S_2,..S_j..]$, and of $B_*$, $[B_1,B_2,..B_j..]$ where each $n_*$ and $S_*$ refer to size $B_*$ of the $p_*^{th}$ category.
We first consider the case where these are all drawn from the same ecosystem and observed on the same lengthscale so that we have the same $\alpha, P, L, R, G$ and $ r^*$. There will be a single function
 $\Psi(p_*)$ which corresponds to this set of observations at different $p_*$.
We can write
\begin{equation}
\Psi(n_*S_*B_*)=\frac{n'S'B'}{n_*S_*B_*}\Psi(n'S'B')
\end{equation}
so that relative to any particular category $n'S'B'$ we can obtain $\Psi(n_*S_*B_*)$ to within a constant.

 This procedure is illustrated in Figure 4
where we have modelled synthetic data.  Our synthetic data are generated such that there are power law  trends in
 abundance, diversity and body size,
 and also random scatter in all variables. Importantly this random scatter is generated to be constrained such that all the sampled categories of individuals share the same function $\Psi$. The plots show that the functional dependence of  $\Psi$ only emerges in a  plot of $1/S_*n_*B_*$ versus body size $B_*$. This also offers a method to compare different categorization schemes which for the same ecosystem could yield different $\Psi(p)$; one could then in principle normalize for (ie compensate for) any 'bias' introduced by a particular choice of categorization.

\end{document}